# Conceptualization of Electromagnetic Induction at various Educational Levels: a Case Study

Michela Cavinato, Elia Giliberti and Marco Giliberti
*Dipartimento di Fisica, Università degli Studi di Milano, via Celoria 16, 20133 Milano, Italy*

Corresponding author: Marco Giliberti
Email for correspondence: marco.giliberti@unimi.it
Phone number: +39 02 50317685

**Abstract:** A vast scientific literature in physics education documents a general widespread difficulty in dealing with Electro-Magnetic Induction (EMI) at various levels of instruction. But, at the best of our knowledge, there is a lack of research that compares difficulties about EMI at different educational degrees. We discuss here a case study about Italian high school, graduate students' and teacher's conceptualization of some aspects of EMI as a function of the sample instruction level. We analyse the answers to a multiple choice written questionnaire, adapted from the literature and given to a total of 49 students. Some difficulties, emerged during the exams of university students of a physics education course while discussing their final project concerning a didactical path about EMI for secondary school, are also discussed. We find that some deep misunderstandings are common at all levels of education and probably come from the very poor link, generally presented in teaching EMI, between the Faraday's flux law and the Lorentz force.

## 1. Introduction

Although the Ørsted experiment, dating back to 1820, is generally seen as the act of birth of the electromagnetic theory, the study of the static electric and magnetic fields can be simply considered nothing more than a great enrichment of the Classical Mechanics viewpoint. It is with the Electro-Magnetic Induction (EMI) that one really enters into the new physics of the Electromagnetic phenomena. This consideration explains the efforts made in recent years (see [1-6] and references therein) by various Physics Education Research (PER) groups to analyse all the problems related to the teaching/learning of EMI. Although EMI is relevant in everyday life (transformers, induction cookers, electric motors and AC power suppliers can be found anywhere), it is nonetheless a phenomenon somewhat far from everyday common experience. Moreover, we think that at school EMI is not faced with enough details, both from an educational and from a practical knowledge perspective. Our opinion, instead, is that EMI is the bridge between an old-fashioned static physics and a much more dynamic modern viewpoint where the fields, playing a central role, are strictly interconnected. In addition, due to the abstractness of the content, misunderstandings of EMI cannot arise only from the simple use of common sense schemes, but also from disciplinary schemes that are not well understood. This fact makes EMI even more interesting from a teaching point of view.

Being interested in a comparative study about the comprehension of EMI at various educational levels, we prepared a written questionnaire that incorporates questions similar to some of those proposed by the literature [1-5] and gave it to university and scientific high school students and physics teachers. In the following, in section 2 we will describe the context and the background while in section 3 we give the research design and method. In section 4 we will analyse and argue the answers to the proposed questionnaire, giving also the percentage of correct answers with sufficiently correct motivations. Some of the difficulties emerged during the exams to university students of a physics education course while discussing their final project concerning a didactical path about EMI for secondary school are reported in section 5. Finally, in section 6, our conclusions are drawn.

## 2. Context and background

The Faraday's flux law for the induced electromotive force (*emf*) reads:

$$emf = -\frac{d\phi(\vec{B})}{dt}, \qquad (1)$$

where $\phi(\vec{B})$ represents the flux of the magnetic field $\vec{B}$. Equation (1) often generates doubts also in physics researchers and university professors. In fact, it is difficult to properly understand EMI in quite a number of different phenomena; for instance in discussing the Faraday disk, the homo-polar generator, the Blondel experiment and similar related devices [1, 6, 7].

Since the middle of the 20th century, there have been many disciplinary discussions on the Faraday's flux law (see [5] and references therein). Its validity has been thoroughly debated from a disciplinary point of view since many authors did not recognize its general applicability so that, as strange as it may seem, only in the last two or three decades most of the experts – with few exceptions (see for instance [6]) – believe that there are no exceptions to Faraday's flux law ([1, 2, 5] and references therein).

If "experts" are confused, the fact that even teachers and students show some uncertainties in dealing with EMI phenomena could be expected. PER groups have actually put in evidence many difficulties coming from the teaching and learning of EMI and there are now quite a few studies about the reasoning of students on EMI and their understanding of it; most (but not all) of them at university level [1, 2, 4-6, 8-10]. We can summarize some of the most common students' difficulties as:

1. Difficulties with the general meaning of flux of a vector field;
2. Confusion between the magnetic field $\vec{B}$ and its time variation;
3. Confusion between the magnetic field $\vec{B}$ and its flux;
4. Misunderstanding about the integration area in the flux rule;
5. Difficulties in understanding the nature of the forces acting on the charges in motion;
6. Difficulty recognizing EMI when an induced electric current is not present.

The situation of the Italian upper secondary school is in some sense peculiar. With respect to most of European and American upper secondary schools, the level of mathematics (that includes elementary calculus) in the Italian upper secondary school curriculum is "high", as is the required disciplinary preparation of the physics teachers, since they must have attained a master degree in physics or in mathematics. Moreover, the recent upgrading of the Italian curriculum requires a necessary and in-depth reflection. As regards more specifically EMI, a document (dated 2015) of the Italian Ministry of Education, University and Research addressed to the scientific high school, indicates knowledge, skills and competences which may be subject to verification in the final written exam. Here a short excerpt [11]:

1. Discuss the physical meaning of the formal aspects of the Faraday-Neumann-Lenz law;
2. Describe, even formally, the relationship between the Lorentz force and the induced electromotive force;
3. Calculate the variations of the magnetic field flux;
4. Using the Faraday-Neumann-Lenz law also in differential form, calculate induced currents and electromotive forces;
5. Solve exercises and problems of application of the studied formulas including those that require the calculation of the forces on moving conductors in a field.

As a consequence, the final exam based on the national curriculum demands knowledge/skills/competences to solve complicated exercises and needs also a robust qualitative framework. These requests, however, seem excessive when compared with the results, reported in the literature, of extensive researches concerning the students' understanding of these topics. Two issues thus quite naturally emerge: are the Italian high school students really in a better situation than the reported international ones? Are Italian high school teachers really prepared to teach EMI in detail?

**3. Research design and method**
Two research questions (RQs) were then formulated.
RQ1) Are the difficulties about the teaching/learning of EMI the same in Italy as those reported in the international literature?
RQ2) If and how, does the conceptual understanding of EMI change at different levels of education, from secondary school students to students attending a master degree in physics/mathematics, to physics teachers?

Our analysis has been performed with the aid of:

1. A 14-questions multiple choice questionnaire given to 16 students of the university course "Preparations of Didactical Experiences" attended, in the academic year 2017-2018 at the University of Milano, by 4 students of the 1st year of the Master Degree in Mathematics (1MDM), 9 students of the 1st year of the Master Degree in Physics (1MDP), and 3 Physics Teachers (PT) who had already taken a Master Degree in Physics. Explanations have always been asked for every question. The questionnaire was delivered in January 2018;
2. A 6-questions multiple choice questionnaire – subset of the previous 14-questions questionnaire – (see Appendix for the English version) given, in the Spring 2018, to 33 students of the last year of a Scientific High School (SHS), attending the Scientific High School "Palli" in Casale Monferrato, a little town about a hundred kilometres west of Milano, that had already faced EMI at school. Explanatory answers have always been asked for every question;
3. Oral exams to university students on a didactical path about EMI for secondary school.

In order to answer correctly, students had to know (and, having in mind the ministerial indications given above, we did expect them to) that an electromotive force is produced by the flux of the time variation of a magnetic field and by the circulation of the magnetic part of the Lorentz force per unit charge, as given by the following formula [1,5]:

$$emf = -\frac{d\phi_{S_\gamma}(\vec{B})}{dt} = -\phi_{S_\gamma}\left(\frac{d\vec{B}}{dt}\right) + C_\gamma(\vec{v} \times \vec{B}), \qquad (2)$$

where we have indicated with $\phi_{S_\gamma}$ the flux through the surface $S_\gamma$ having $\gamma$ as boundary and with $\vec{v}$ the velocity of the part of $\gamma$ subjected to the magnetic field $\vec{B}$.

**4. Results from the 14-questions and the 6-questions multiple choice questionnaires**
In order to make a comparative analysis of the understanding of EMI at different levels of education, from the 14-questions multiple choice questionnaire (point 1. of section 3) we extract only the six questions proposed in the 6-questions multiple choice questionnaire (point 2. of section 3). Questions D1, D2 and D3 are essentially questions 29, 30 and 31 of the area X (Faraday's law) of the "Conceptual Survey in Electricity and Magnetism" [3]; questions D4, D5 and D6 are similar to questions Q4 of [2], Q1 and Q3 of [5].
At first glance, questions D4 - D6 could appear difficult, especially for high school students, and even university professors might be challenged by them. In our opinion, they are not really difficult, but only appear to be so, principally because they are not usually presented in traditional textbooks. The reason for this absence is not related to their difficulty – questions are only qualitative and calculations are absent – but to the fact that the last term in equation (2) is often not well presented and discussed. For each question, the explanations have been grouped into categories according to the type of reasoning used, independently of the correctness of the answer. The categorization has been made through a "negotiation phase", starting from the individual analysis of the answers made by everyone of the authors. Therefore the categories were not given *a priori* and have been constructed to be mutually exclusive. The answers without explanation have been put into the general set "No expl".
In the following, we give, for each of the 6 questions, a table with our categorization and some examples of the explanations given by the test participants, together with few comments. For completeness, for questions D1-D3, to the explanations presented as examples we also add the chosen answers, in square brackets and before the reported sentence.
To make the tables easier to read, let us briefly remember the acronyms used:

- SHS: students of the last year of a Scientific High School;
- 1MDM: students of the 1st year of the Master Degree in Mathematics;

- 1MDP: students of the 1st year of the Master Degree in Physics;
- PT: Physics Teachers already holding a Master Degree in Physics.

Finally, in subsection 4.7, not only for completeness, but also to better illustrate the quality of the answers and their explanations, we provide the percentage of students who have both put the cross on the right answer in the multiple choice questionnaire and given sufficiently correct arguments to motivate that choice.

*4.1. Question D1 (Fig. A1 in Appendix)*
Question D1 concerns the current that is possibly induced in a circuit when the magnetic field is variable or the loop is moving. Table 1 summarizes our categorization giving the number of explanations per each category (in the rows) of each coherent students group (in the columns).

| Table 1. Answers to D1 | | | | |
|---|---|---|---|---|
| **D1** | **SHS** (33) | **1MDM** (4) | **1MDP** (9) | **PT** (3) |
| **Flux** | 16 | 1 | 8 | 0 |
| **B** | 2 | 2 | 1 | 2 |
| **Confusion between primary and self-induced B** | 1 | 0 | 0 | 0 |
| **No expl** | 14 | 1 | 0 | 1 |

Here and in the following, the "Flux" category groups the explanations in which the cause of EMI is principally ascribed to the variation of the magnetic field flux. For instance:
SHS: [c] "*In case I, the magnet moves away creating a variation in the flux of the magnetic field that generates an induced current.*".
SHS: [c] "*An induced current is provided by a change of flux. In the first and fourth cases, either by moving the magnet away from or by approaching the coil, there is a variation of the magnetic field and, consequently, also a flux variation which leads to the creation of an electric current. In the second case the radius of the loop decreases and therefore there is a variation of the surface.*".
1MDM: [c] "*I. The flux varies because the magnet moves away and then the magnetic field of the loop changes; II. The flux varies because the surface of the loop decreases and the field is fixed; IV. It is similar to I, but it is the wire which is approaching.*".
1MDP: [c] "*I must have a flux variation through the loop. In I and IV, I have it for the different field intensity; in II, I have it because the surface changes; in III, I do not have it because neither the field nor the surface change.*".
In the " B " category, here and in the following, we put the explanations in which the cause of EMI is to be ascribed to the time variation of the magnetic field. Examples are:
SHS: [c] "*In the first case, the magnet, moving away from the loop, creates a variation of magnetic field that acts on the loop, since the field lines become weaker.*".
SHS: [b] "*I: the fact that the magnet is moving generates a variation of the magnetic field which causes an electric current to be generated.*".
1MDM: [c] "*The magnet that moves away or approaches changes the magnetic field inside the loop. Even the tightening wire changes the magnetic field inside the loop [...]*".
1MDP: [b] "*Case I and case IV are the only ones in which there is a variation of quantities to induce a variation of the field B which induces a current.*".
PT: [c] "*In the cases reported, I have a variation of magnetic field concatenated to the loop. The induced current is a current that is generated in the loop to oppose the variation of the magnetic field.*".
We also found one high school student that made "Confusion between the primary and the self-induced B ". In fact he wrote that:
SHS: [c] "*In I and IV, there is a variation of the magnetic flux, which induces in the coil a magnetic field, which induces a current.*".

As a comment, we see that nearly 90% of the 1st year Master Degree students in Physics and 50% of the Scientific High School students reasoned in terms of flux variation, while no Physics Teacher did this way. Moreover, we want to emphasize that nobody gave an explanation in terms of the magnetic part of the Lorentz force that is, instead, the cause of EMI in the cases II and IV.

*4.2. Question D2 (Fig. A2 in Appendix)*

Question D2 pertains the current that is possibly induced in a circuit when it is nearby a very long current carrying wire. Table 2 summarizes our categorization.

| Table 2. Answers to D2 | | | | |
|---|---|---|---|---|
| **D2** | **SHS** (33) | **1MDM** (4) | **1MDP** (9) | **PT** (3) |
| **Flux** | 10 | 1 | 6 | 1 |
| **B** | 3 | 2 | 1 | 0 |
| **Current-velocity angle** | 11 | 0 | 0 | 0 |
| **No expl** | 9 | 1 | 2 | 2 |

Examples of the "Flux" category are:
SHS: [a] "*The magnetic field lines generated by the current i in the wire are thinning going away from the wire. Therefore, only in the case that the displacement has a component perpendicular to the wire, there will be a variation of the flux of the magnetic field and therefore an induced current.*".
1MDM: [a] "*The magnetic field generated by the wire depends on r. In order for a change in the flow to occur in the (rigid) loop, the loop will have to move away or approach the wire, while maintaining a parallel motion there is no flux, no induced current.*".
1MDP: [a] "*There is induced current if there is a variation of flux through the surface of the loop. […] Case I: the flux changes because the intensity of the field decreases; Case II: the same as in the case I; Case III: I have no change of flux because the loop is always at the same distance from the wire.*".
PT: [a] "*In the case of variable B, since $\nabla \cdot B = 0$ the flux of the magnetic field lines through a closed surface must be zero, this is possible only if an induced B is generated.*".
Three answers coming from the " B " category are:
SHS: [d] "*Whether it moves in one direction or the other, what changes is the intensity of the field that acts on it.*".
1MDM: [c] "*The correct answer is c because, in the case I, the magnetic field is always perpendicular and therefore does not induce a current.*".
1MDP: [d] "*The magnetic field of an infinite wire is on circular lines of force. In all the three cases the velocity is orthogonal to the magnetic field so there is always an induced electric field, in all the cases.*".
We also found that one third of the SHS (but no one of the others) think EMI is in some way related to the angle formed between the current direction and the velocity of the loop ("Current-velocity angle" category). For instance:
SHS: [c] "*In case I, the angle between the wire and the moving coil is 90°; the cosine is null and therefore the flux is null and there is no electromagnetic induction.*".
SHS: [c] "*The loop will be crossed by current only when the current and the velocity of the loop are not placed at 90° because the change of flux is nullified.*".

*4.3. Question D3 (Fig. A3 in Appendix)*
The presence of EMI in the absence of a closed loop is taken into account in question D3.

| Table 3. Answers to D3 | | | | |
|---|---|---|---|---|
| **D3** | **SHS** (33) | **1MDM** (4) | **1MDP** (9) | **PT** (3) |
| **Flux** | 9 | 0 | 1 | 1 |
| **Lorentz** | 4 | 1 | 2 | 0 |
| **No separation because the bar is neutral** | 14 | 0 | 1 | 1 |
| **No expl** | 6 | 3 | 5 | 1 |

Examples of explanations from the "Flux" category are:
SHS: [e] "*According to the right hand rule, the flux coming out of the plane, creates a current that moves counter clockwise.*".
1MDP: [a] "*There is no variation in the flux of the magnetic field (the surface of the bar hit by the field is always the same) so there is no induced current.*".
In the "Lorentz" category, here and in the following, we put the explanations in which the cause of EMI is to be ascribed to the magnetic part of the Lorentz force. Here below, some explanations coming from this category:
SHS: [e] "*By the right hand rule, the force created on the bar is directed downwards, so the distribution of the charges that is formed is that of the case (e).*".
1MDM: [d] "*By the Lorentz force the charges are pushed downward.*".
1MDP: [e] "*A Lorentz force is acting on the charges of the bar. The positive charges then move downwards and the negative ones upward (towards the opposite side of the force).*".
In the "No separation because the bar is neutral" category (we believe that the name of this category is self-explanatory) we find (for example):
SHS: [a] "*The motion of the bar at velocity v in a constant and uniform magnetic field B does not change the charges of the neutral bar; therefore it remains the same as the initial figure.*".
We point out that about 43% of the SHS students fall in this category.

*4.4. Question D4 (Fig. A4 in Appendix)*
Question D4 concerns the current induced in a loop rotating in a uniform magnetic field and asks for the nature of the microscopic forces in action.

| **Table 4. Answers to D4** | | | | |
|---|---|---|---|---|
| **D4** | **SHS** (33) | **1MDM** (4) | **1MDP** (9) | **PT** (3) |
| **Flux** | 25 | 1 | 6 | 0 |
| **B** | 0 | 1 | 1 | 2 |
| **Lorentz** | 0 | 0 | 1 | 1 |
| **No expl** | 8 | 2 | 1 | 0 |

It is interesting to observe that the vast majority of students belongs to the "Flux" category and that only 2 (one 1MDP and one PT) among the 49 students were able to explain that the force acting on the moving charges is the (local) magnetic part of the Lorentz force ("Lorentz" category). In the following the 1MDP and the PT explanations:
1MDP: "*[…] is the Lorentz force on moving electrons. This is due to the fact that there is a vector product in the Lorentz force.*".
PT: "*The conduction electrons, free to move inside the conductor, move with velocity v = v (therm) + v (cond) where v (therm) is the velocity due to the thermal agitation and v (cond) the velocity of their part of the conductor at a given time. The velocity v (therm) is distributed in all directions while the velocity of the rotating conductor is not. Therefore the electrons are affected by the Lorentz force […]*".

*4.5. Question D5 (Fig. A5 in Appendix)*
Question D5 is about a C-shaped conductive wire sliding along a conductive magnet. In this question the importance of equation (2) is particularly evident. In fact, an electromotive force is generated in the wire by the Lorentz force magnetic term. On the contrary, nearly all the explanations, whether coming from the High School students, from the Master Degree students or from the Physics

Teachers, are biased by the common belief that, since when the loop is stationary there is no flux because the magnetic field lines are parallel to the loop surface, there is still no flux (variation) when the loop is moving (see the "No current because no flux variation" category).

| Table 5. Answers to D5 | | | | |
|---|---|---|---|---|
| **D5** | **SHS** (33) | **1MDM** (4) | **1MDP** (9) | **PT** (3) |
| **No current because no flux variation** | 28 | 0 | 3 | 2 |
| **Lorentz** | 0 | 0 | 2 | 0 |
| **Flux cutting** | 1 | 0 | 0 | 0 |
| **No expl** | 4 | 4 | 4 | 1 |

From the "No current because no flux variation" category, we can report:
SHS: "*The magnetic field lines form an angle of 90° with the normal to the surface identified by the wire and part of the magnet. This means that the flux is zero and persists in its value as the wire slides downwards. Therefore, there will be no induced current.*".
1MDP: "*There is no induced current because there is no variation of the flux of $\vec{B}$ (the angle is always the same).*".
PT: "*No current, the lines concatenated to the loop do not vary.*".
From the "Lorentz" category:
1MDP: "*The downward movement of the wire implies the vertical movement of the charges inside it and therefore a current.*".
It is particularly interesting to observe that one SHS, with no previous knowledge of flux cutting, spontaneously described/noticed a variation of the area ("Flux cutting" category):
SHS: "*The area changes because the wire moves up and down.*".

### *4.6. Question D6 (Fig. A6 in Appendix)*
Question D6 concerns the presence or absence of EMI when in a uniform and time-independent magnetic field a switch closes a circuit and opens a wider one (see also [1]; a question very close to this one has been previously proposed by Tilley in [12]). The common *naïve* reasoning that there is a flux variation because the area of the circuit changes dominates the answers and spans all the groups of students. We also found 6 SHS reasoning in terms of a not well specified self-inductance of the circuit.

| Table 6. Answers to D6 | | | | |
|---|---|---|---|---|
| **D6** | **SHS** (33) | **1MDM** (4) | **1MDP** (9) | **PT** (3) |
| **Area/Flux** | 8 | 1 | 5 | 1 |
| **Open circuit** | 6 | 0 | 1 | 0 |
| **B changes since the area changes** | 0 | 1 | 0 | 0 |
| **Self-inductance** | 6 | 0 | 0 | 0 |
| **No expl** | 13 | 2 | 3 | 2 |

Examples of "Area/Flux" category:
SHS: "*An induced current circulates because, passing from position A to position C, there is an increase in the area of the circuit immersed in the magnetic field. Varying the area also the flux varies.*".
1MDP: "*Yes, moving from position A to position C the circuit increases its area. Since the magnetic field is uniform, the flux increases instantaneously and there is an induced current spike.*".
In the "Open circuit" category we can find:

SHS: "*The fact that the switch disconnects from A and closes the circuit in C does not allow the formation of an induced current because in the intermediate passage the circuit remains open.*".
1MDP: "*While the switch goes from A to C the wire is interrupted so a current cannot pass.*".
Concerning the "Self-inductance" category we reads, for examples:
SHS: "*For a short time, we will observe the passage of current which will lead to the variation of the magnetic flux and to the creation of an electromotive force and therefore of an induced current in the opposite direction.*".

*4.7. Percentage of correct answers and explanations to the questionnaire*
As written above, we give here the percentage of students who gave the correct answer in the multiple choice questionnaire together with a sufficiently correct motivation. Given the small number of elements of some groups of students (for example, the teachers were only three) we do not consider it appropriate to provide the percentages of correct answers and explanations by type of students.
For what concerns question D1, we observe that although nobody gave an explanation in terms of the magnetic part of the Lorentz force (that is the cause of EMI in the cases II and IV), since the question only asks "In what situation will the coil be crossed by an induced current?" and not the nature of the forces acting on the charges, we assume as correct all the explanations that ascribe to a generic flux variation the presence of an induced current. On the contrary, in the question D4, that reads "Explain the origin of the forces that move the charges in the loop.", we considered wrong the generic non local explanations given in terms of flux variation.

1. Percentage of correct answers + explanation to D1: 47%
2. Percentage of correct answers + explanation to D2: 39%
3. Percentage of correct answers + explanation to D3: 16%
4. Percentage of correct answers + explanation to D4: 4%
5. Percentage of correct answers + explanation to D5: 4%
6. Percentage of correct answers + explanation to D6: 6%

**5. Oral exams**
As a final exam of the university course "Preparations of Didactical Experiences" held by one of us (MG) - in which a discussion about the teaching of EMI has been performed with an analysis of equation (2) - a group of five students (four of 1MDM and one of 1MDP, all attending the course in the academic year 2018-2019), that must not be confused with the ones answering our questionnaire, elaborated a didactic path about EMI addressed to scientific high school. During the exams we were principally interested in two conceptual aspects:

1. Which is the physical origin of the force that moves the charges in an EMI process?
2. Which is the link between the magnetic part of the Lorentz force and the Faraday's flux law?

To that purpose, we asked the students to discuss these points making always reference to the six questions analysed in section 4 (of which only the question D4 faced the aspect 1. and no questions faced the aspect 2.).
When in presence of a moving conductor in a constant magnetic field, all students were confident to identify the force acting on the charges with the magnetic part of the Lorentz force. More difficulties emerged when dealing with a still conductor in a time-dependent magnetic field. An excerpt of the discussion can help understanding.
Concerning case I of question D1.
Professor: "*Since the coil is crossed by an induced current, which force is acting on each charge?*".
Student 1: "*The electric part of the Lorentz force.*".
Professor: "*And what is this electric field created from?*".
At this point all the students seemed perplexed.

Student 2 (the one studying physics): "*The variation of the magnetic field.*".
Professor: "*Do you mean you know a formula in which you can write $\vec{E}$ equal to something containing the time derivative of $\vec{B}$?*".
Student 2 was again perplexed and gave no answer.
Although we are well aware that most introductory university textbooks on electromagnetism ascribe the presence of an induced electric field to the generic time variation of a magnetic field, we believe that the knowledge of the Jefimenko equations [13-15] and/or of an explicit formula that gives the local direct expression of the electric field in terms of the variations of the (electric scalar and magnetic vector) potentials [6, 15, 16], would deeply help even a qualitative understanding of EMI.
Even in the case of the aspect 2. an excerpt of the discussion can be useful.
Concerning question D3.
Professor: "*With the help of the Lorentz force, you have all answered the question [D3] correctly. How could you answer using the Faraday's flux law?*".
Student 2: "*The bar is not a closed circuit, therefore I cannot use the flux law.*".
Student 3: "*I read in a paper that the flux law has no general validity...*".
Professor: "*Which paper?*".
Student 3: "*I cannot remember it now.*".
As already widely observed in the literature [1, 5], students have great difficulties in finding the surface for the calculation of the magnetic flux in the case of a moving conductor.
We observe that, at variance with the results reported in section 4, these five students we examined – probably for having attended a course in physics education and having deepened their study on EMI, reading specialized literature and elaborating a didactical path for high school – show nearly no difficulties in dealing with motional electromotive force.

## 6. Conclusions

For what concerns RQ1, the majority of our results are in agreement with those found in the literature [1, 2, 4-6, 8-10]. In particular, most of the students use the Faraday's flux law even when a description in terms of the magnetic part of the Lorentz force could be more appropriate. When using the flux law, students are in general unable to consider the correct surface for the magnetic flux calculation, both in the presence and in the absence of the circuit motion. Most students are unable to describe the induced current considering the local forces acting on the charge carriers (see D4).
As one might expect, questions D1 - D3 are those which show the highest percentage of correct answers and explanations, since they are much more in line with standard presentations; nonetheless less than half of the sample group was able to explain even these questions correctly. This is probably a symptom of the fact that the comprehension of EMI is far from being acceptable.
Regarding RQ2, that is at the core of our somewhat novel study, we observe that the level of understanding remains substantially unchanged with growing the level of education. For example, the greatly preferred use of the Faraday's law with respect to the Lorentz force in dealing with EMI (in general confusing the circuit surface with the integration area) and the ability to connect a systemic description (electromotive force) to locally acting forces do not appreciably change going from high school students to physics teachers. From a didactical point of view, we believe that some more attention should be deserved to this last point, since a deeper comprehension of the locally acting forces may lead to a better understanding of the physical roots of EMI. In fact, while it is surely the magnetic part of the Lorentz force the force acting on a conducting wire moving in a constant magnetic field, on the contrary, it is not the magnetic part of the Lorentz force the one acting on the wire when it stays still and the source of the non-uniform magnetic field is moving. This fact may be very confusing (consider, for example, we are interested in the forces acting in the situations described in D1). The asymmetry found between the physical explanation of the two situations (conductor in motion - sources of $\vec{B}$ fixed; conductor fixed - sources of $\vec{B}$ in motion) is well known, but it has its full resolution only in special relativity. Nonetheless, especially for students that do not know relativity, if we do not linger upon this asymmetry problem, the common explanation of the force acting locally on the wire when the sources are in motion is that it is due to the non-conservative electric field generated by the time-varying magnetic field. On the other hand, the Maxwell equation $\nabla \times \vec{E} = -\frac{\partial \vec{B}}{\partial t}$, which one often refers to, gives only a sort of topological (it gives the curl of $\vec{E}$ ) local property of the electric field and means that, if there is a time-dependent magnetic field, there is also, simultaneously, a non Coulombian electric field; however it does not give $\vec{E}$ explicitly (see section 5).

The difficulties we found in students' understanding of EMI, persistent at all educational levels, indicate that the traditional, somewhat standard introduction of EMI via the Faraday's flux law, evidently presents some educational bugs. From our results, it also follows that a way to better understand EMI should pass through a more strong connection between the flux law and the Lorentz force. The results of the experimentation of a new didactical path proposed in the last year of a scientific high school in northern Italy will be presented and analysed in a forthcoming paper.

**Acknowledgments**
We thank Marco Stellato for the useful discussions.

**Appendix**
The four figures below represent a cylindrical magnet and a small copper coil lying on a plane perpendicular to the NS axis of the magnet. The motion states of the magnet and the coil are shown ($\vec{v}$ represents the velocity). In what situation will the coil be crossed by an induced current? Motivate your answer.

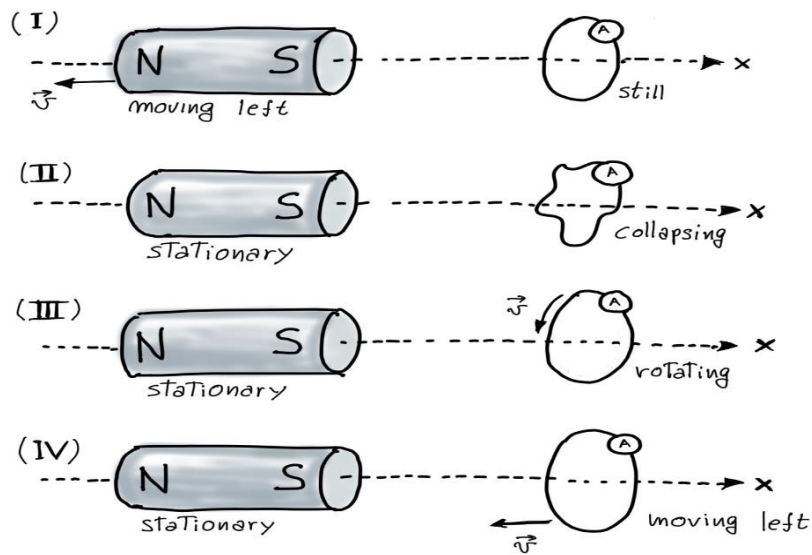

(a) I, III, IV  –  (b) I, IV  –  (c) I, II, IV  –  (d) IV  –  (e) No one

Figure A1: QUESTION D1

A very long rectilinear wire carries a DC current i. A rectangular metallic loop moves with velocity $\vec{v}$ in the same plane of the wire, in the direction indicated in the three figures. In which cases will the coil be crossed by an induced current? Motivate your answer.

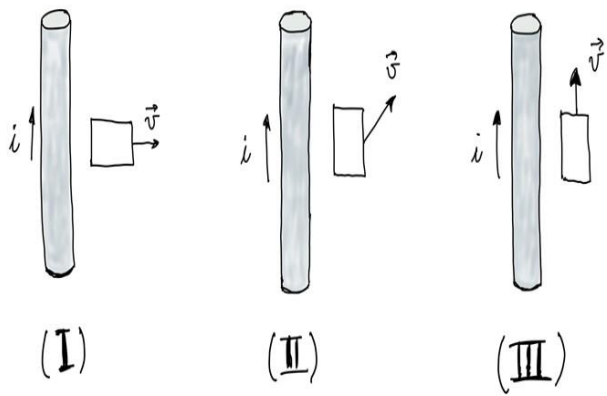

(a) Only I and II – (b) Only I and III – (c) Only II and III – (d) In all the cases – (e) In none of the above cases.

Figure A2: QUESTION D2

A neutral metal bar moves at a constant velocity $\vec{v}$ in a uniform and stationary magnetic field with direction into the page (and therefore perpendicular to the velocity of the bar). Which of the diagrams at the bottom best describes the charge distribution on the surface of the metal bar? Motivate your answer.

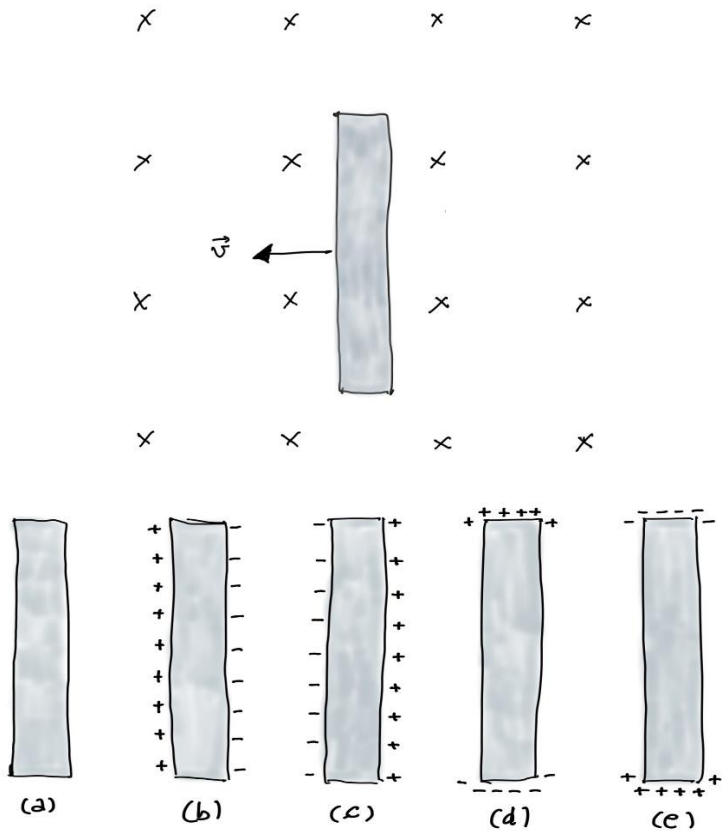

Figure A3: QUESTION D3

A circular conductive loop, placed in a steady, uniform, vertical magnetic field, is connected to an ammeter. When the loop is rotated upwards, the ammeter measures a current. Explain the origin of the forces that move the charges in the loop. Motivate your answer.

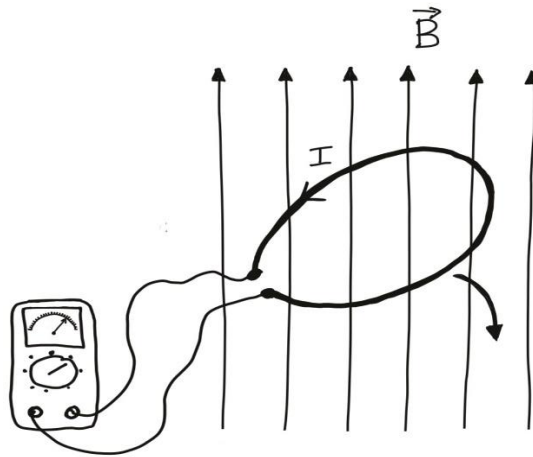

Figure A4: QUESTION D4

A C-shaped wire is slipping inside the air gap of a magnet, keeping the contact with one of the poles and remaining constantly on a plane parallel to the magnetic field inside the air gap. Keeping in mind that both the wire and the magnet are conductors, determine whether an induced current flows in the wire. Motivate your answer.

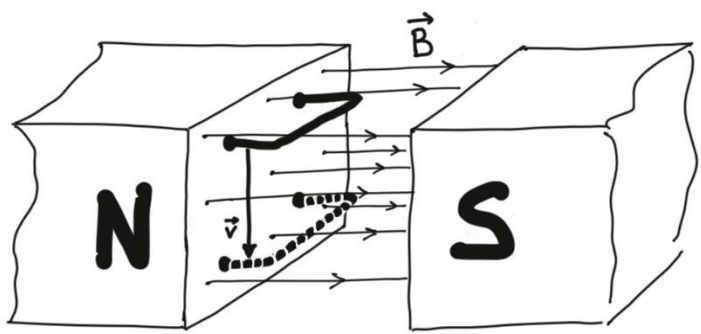

Figure A5: QUESTION D5

The figure below shows a conductive circuit in a steady and uniform magnetic field coming out of the sheet. Determine whether an induced current will flow in the circuit when the switch passes from position A to position C. Motivate your answer.

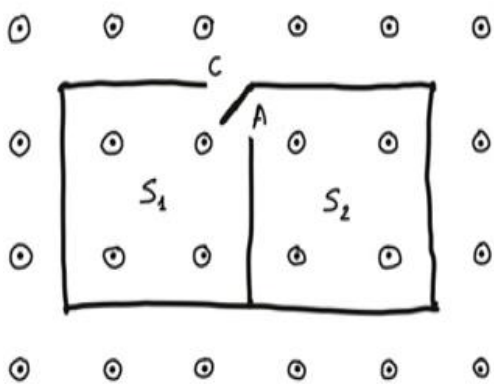

Figure A6: QUESTION D6